\documentclass[aps,pre,reprint,superscriptaddress,amsmath,amssymb]{revtex4-2}
\usepackage{graphicx}
\usepackage{bm}
\usepackage{xcolor}
\usepackage[colorlinks=true,linkcolor=blue,citecolor=blue,urlcolor=blue]{hyperref}

\newcommand{\mct}{\mathrm{iMCT}}
\newcommand{\tmct}{\mathrm{iTMCT}}
\newcommand{\dd}{\mathrm{d}}

\begin{document}

\title{Exponent parameter and intrinsic wave-vector cutoff in the ideal
time-convolutionless mode-coupling theory of hard spheres}

\author{Yuto Kimura}
\email{kimura-y@t.kagawa-nct.ac.jp}
\affiliation{Department of Mechanical Engineering, National Institute of Technology (KOSEN),
Kagawa College, 355 Chokushi-cho, Takamatsu, Kagawa 761-8058, Japan}

\date{\today}

\begin{abstract}
The exponent parameter $\lambda$ of mode-coupling-type theories fixes all asymptotic
exponents of glassy dynamics near the critical point and is routinely inferred from fits to
simulation and experimental data. We determine $\lambda$ for the ideal time-convolutionless
mode-coupling theory (ideal TMCT), which predicts a critical volume fraction of hard spheres
close to simulation values, and compare it with ideal mode-coupling theory (ideal MCT) for
Percus--Yevick hard spheres without damping. Deriving the $\beta$-scaling equation of ideal
TMCT, we show that $\lambda$ consists of an MCT-type term and an additional positive term,
half a weighted average of the critical cumulant $K^c=-\ln f^{\infty,c}$, which vanishes only
when $f^{\infty,c}\to1$. For hard spheres this gives $\lambda=0.855$, compared with 0.707 for
ideal MCT, a value confirmed by four independent methods. Ideal TMCT has an intrinsic cutoff
$q^*\sigma\simeq28$, above which $\varphi_c=0.5824$ and $\lambda$ no longer depend on the
wave-vector cutoff; below $q^*$ the exponents are far more sensitive to the cutoff than
$\varphi_c$. Fits near the main peak of the structure factor nevertheless yield an apparent
$\lambda\simeq0.735$, close to the ideal MCT value, and in the range accessible to
simulations the effective exponent of the self-diffusion coefficient $D_s$ is 2.6 rather than
its asymptotic value 3.35. Without damping, $D_s$ depends logarithmically on the cutoff in
both theories, a consequence of the contact discontinuity of the hard-sphere direct
correlation function rather than of the Percus--Yevick approximation.
\end{abstract}

\maketitle

\section{Introduction}

The ideal mode-coupling theory (ideal MCT) of the glass transition
\cite{Bengtzelius1984,Leutheusser1984,GotzeBook} predicts an ergodic-to-nonergodic
transition at a critical point and universal asymptotic laws for the two-step relaxation
near it \cite{Gotze1991}. For hard spheres with the Percus--Yevick (PY) structure factor
\cite{PercusYevick1958}, ideal MCT gives the critical volume fraction $\varphi_c\simeq0.516$
\cite{Franosch1997,Chong2001,Voigtmann2004}, well below the values
$\varphi_c\simeq0.58$--$0.59$ obtained from simulations and colloidal experiments
\cite{Tokuyama2003,Tokuyama2007,Weysser2010,Brambilla2009}. One line of work extends ideal
MCT itself. In generalized MCT, the four-point correlation functions
that ideal MCT factorizes are treated by their own equations of motion, and the
factorization is applied only at a higher level
\cite{Szamel2003,WuCao2005,JanssenReichman2015}. For PY hard spheres the resulting $\varphi_c$
increases with the closure level and reaches about 0.56 at the sixth level
\cite{LuoJanssen2020}.

A different approach was taken by Tokuyama, who proposed the ideal time-convolutionless
mode-coupling theory (ideal TMCT) \cite{Tokuyama2014,Tokuyama2015}. It uses the same mode-coupling approximation for
the memory function as ideal MCT, but treats the density with the time-convolutionless projection
operator method \cite{TokuyamaMori1975,TokuyamaMori1976} instead of the convolution type
\cite{Mori1965}. The resulting equation for the cumulant $K=-\ln f$ of the scaled
intermediate scattering function $f$ contains no time convolution of $f$ itself. The full
time-convolutionless mode-coupling theory, now referred to simply as TMCT and called full TMCT in this paper, was developed later
\cite{Tokuyama2017}; ideal TMCT follows from it by a Markov approximation for the memory term,
and the two theories have the same critical point \cite{Tokuyama2026}. For PY hard spheres
ideal TMCT gives $\varphi_c\simeq0.582$ \cite{KimuraTokuyama2016,NarumiTokuyama2017}, close
to the simulation values. The theory has also been applied to attractive colloids
\cite{NarumiTokuyama2017} and to the relation between the wave-vector cutoff and a
characteristic length at the $\beta$ stage \cite{TokuyamaNarumi2019}.

Two questions remain open. The first concerns the asymptotic laws. G\"otze and Schilling
analyzed a schematic version of ideal TMCT and found the same scenario as in ideal MCT, with different
anomalous exponents \cite{GotzeSchilling2015}. In contrast, an estimate based on fits of
earlier numerical solutions for hard spheres gave $\lambda\simeq0.733$, close to the ideal MCT
value \cite{Tokuyama2026}. The second question concerns the wave-vector cutoff $q_c$, which
in full TMCT and ideal TMCT is regarded as a physical parameter. Tokuyama and Narumi found that the critical
point of ideal TMCT does not change for $q_c\ge1/b_c$ with $\sigma/b_c=27$, where the
nonergodicity parameter vanishes at $q=1/b_c$ \cite{TokuyamaNarumi2019}. In applications of
full TMCT \cite{Tokuyama2017} and of ideal TMCT \cite{TokuyamaNarumi2019} with damping, the cutoff
has been determined at each volume fraction so that the theory reproduces the simulated $D_s$, and in the supercooled state it lies between the main peak
of $S(q)$ and $1/b_c$ \cite{Tokuyama2017,TokuyamaNarumi2019}. Whether the exponents, unlike
$\varphi_c$, depend on the cutoff in this range has not been examined.

In this paper we address both questions for PY hard spheres within ideal TMCT, with the
damping terms neglected. We derive the $\beta$-scaling equation of ideal TMCT, evaluate
$\lambda$ from the stability matrix, and test it by three independent methods. We show that
ideal TMCT contains an intrinsic cutoff $q^*$ beyond which $\varphi_c$ and $\lambda$ no longer
depend on $q_c$, and that the exponents depend much more strongly on $q_c$ below $q^*$ than
$\varphi_c$ does. Finally we compare the two theories for the mean-square displacement and
the long-time self-diffusion coefficient $D_s$, and discuss the difference between
asymptotic and effective exponents in the comparison with simulations. The improvement of
$\varphi_c$ itself has been reported before \cite{KimuraTokuyama2016,NarumiTokuyama2017} and
is not the subject of this paper.

Section~\ref{sec:theory} summarizes the equations. Section~\ref{sec:beta} derives the
$\beta$-scaling equation and the expression for $\lambda$. Section~\ref{sec:numerics}
describes the numerical method. Results are presented in Sec.~\ref{sec:results} and discussed
in Sec.~\ref{sec:discussion}.

\section{Ideal MCT and ideal TMCT}
\label{sec:theory}

We consider $N$ hard spheres of diameter $\sigma$ and mass $m$ in volume $V$ at temperature
$T$, with number density $\rho=N/V$ and volume fraction $\varphi=\pi\rho\sigma^3/6$. Let
$\bm X_j(t)$ be the position of particle $j$ ($j=1,\dots,N$) at time $t$. The collective and
self density fluctuations are
\begin{subequations}
\label{eq:rho}
\begin{align}
\rho_c(\bm q,t)&=\frac{1}{N^{1/2}}\sum_{j=1}^Ne^{i\bm q\cdot\bm X_j(t)},\\
\rho_s(\bm q,t)&=e^{i\bm q\cdot\bm X_j(t)},
\end{align}
\end{subequations}
for $\bm q\neq\bm0$, where $j$ in $\rho_s$ denotes a tagged particle. The scaled intermediate
scattering functions are
\begin{equation}
f_\alpha(q,t)=\frac{\langle\rho_\alpha(\bm q,t)\rho_\alpha(\bm q,0)^*\rangle}{S_\alpha(q)}
\qquad(\alpha=c,s),
\label{eq:fdef}
\end{equation}
where $\langle\cdots\rangle$ denotes the equilibrium ensemble average, $q=|\bm q|$,
$S_s(q)=1$, and $S_c(q)=S(q)=\langle|\rho_c(\bm q,0)|^2\rangle$ is the static structure
factor. The mean-square displacement of a tagged particle is
$M_2(t)=\langle|\bm X_j(t)-\bm X_j(0)|^2\rangle$. Lengths are
measured in units of $\sigma$, masses in units of $m$, energies in units of $k_BT$, and times in
units of $t_0=\sigma/v_{th}$, where $v_{th}=(k_BT/m)^{1/2}$ is the thermal velocity; wave numbers
are thus given in units of $\sigma^{-1}$ and diffusion coefficients in units of $\sigma v_{th}$.

With the frequency $\Omega_\alpha(q)$ given by $\Omega_\alpha^2(q)=q^2v_{th}^2/S_\alpha(q)$, the ideal MCT
equation reads
\begin{equation}
\ddot f_\alpha(q,t)=-\Omega_\alpha^2\Big[f_\alpha(q,t)
+\int_0^t m_\alpha(q,t-s)\,\dot f_\alpha(q,s)\,\dd s\Big],
\label{eq:mct}
\end{equation}
and the ideal TMCT equation for $K_\alpha(q,t)=-\ln f_\alpha(q,t)$ reads
\cite{Tokuyama2014,Tokuyama2015}
\begin{equation}
\ddot K_\alpha(q,t)=\Omega_\alpha^2\Big[1
-\int_0^t m_\alpha(q,t-s)\,\dot K_\alpha(q,s)\,\dd s\Big].
\label{eq:tmct}
\end{equation}
Equation~\eqref{eq:tmct} follows from full TMCT \cite{Tokuyama2017} by a Markov approximation for
the memory term \cite{Tokuyama2026}. The subscripts iMCT and iTMCT denote ideal MCT and ideal
TMCT, respectively.

Equations~\eqref{eq:mct} and \eqref{eq:tmct} are written without damping terms. In their
general form, both ideal theories contain a damping term on the right-hand side,
$-\gamma_\alpha\dot f_\alpha$ in Eq.~\eqref{eq:mct} and $-\gamma_\alpha\dot K_\alpha$ in
Eq.~\eqref{eq:tmct}, whose damping constant $\gamma_\alpha$ accounts for short-time friction not
contained in the mode-coupling memory function \cite{Tokuyama2017,Tokuyama2026}. In
applications, $\gamma_\alpha$ has been determined from simulation data at short times
\cite{Tokuyama2017,TokuyamaNarumi2019}. Throughout this paper we set $\gamma_\alpha=0$ in both
theories. This is a separate approximation: the word ``ideal'' refers to the Markov
approximation in ideal TMCT and to the neglect of activated hopping processes in ideal MCT
\cite{Gotze1991}. The nonergodicity parameters, the critical point, and $\lambda$ do not
depend on $\gamma_\alpha$, whereas the time scales and the magnitude of $D_s$ do
(Sec.~\ref{sec:discussion}).

The memory kernels are the same in both theories:
\begin{align}
m_c(q,t)&=\frac{\rho S(q)}{2q^4}\!\int\!\frac{\dd^3k}{(2\pi)^3}\,S(k)S(p)
\nonumber\\
&\quad\times\big[\bm q\!\cdot\!\bm k\,c(k)+\bm q\!\cdot\!\bm p\,c(p)\big]^2
f_c(k,t)f_c(p,t),
\label{eq:mc}\\
m_s(q,t)&=\frac{\rho}{q^4}\!\int\!\frac{\dd^3k}{(2\pi)^3}\,S(k)
\big[\bm q\!\cdot\!\bm k\,c(k)\big]^2 f_c(k,t)f_s(p,t),
\label{eq:ms}
\end{align}
with $\bm p=\bm q-\bm k$ and $|\bm k|,|\bm p|<q_c$, where $q_c$ is the wave-vector cutoff.
Here $c(q)$ is the direct correlation function, related to $S(q)$ by the Ornstein--Zernike
relation $\rho c(q)=1-1/S(q)$.

Equations~\eqref{eq:mct} and \eqref{eq:tmct} have a common form. With $y=1-f_\alpha$ for ideal MCT
and $y=K_\alpha$ for ideal TMCT,
\begin{equation}
\ddot y(t)=\Omega_\alpha^2\Big[1-a_\theta\,y(t)-\int_0^t m(t-s)\,\dot y(s)\,\dd s\Big],
\label{eq:unified}
\end{equation}
with $y(0)=\dot y(0)=0$, and $a_\theta=1$ (ideal MCT) or $a_\theta=0$ (ideal TMCT). The two theories
thus differ only by the term $a_\theta y$. For vanishing memory this term gives the undamped
oscillation $f=\cos\Omega t$ in ideal MCT, whereas ideal TMCT gives the Gaussian decay
$f=e^{-\Omega^2t^2/2}$.

In both theories the mean-square displacement obeys \cite{Tokuyama2014}
\begin{equation}
\ddot M_2(t)=6v_{th}^2-\int_0^t\Delta\varphi_s^{(0)}(t-s)\,\dot M_2(s)\,\dd s,
\label{eq:msd}
\end{equation}
\begin{equation}
\Delta\varphi_s^{(0)}(t)=\frac{\rho v_{th}^2}{6\pi^2}\int_0^{q_c}\dd k\,k^4c(k)^2S(k)
f_c(k,t)f_s(k,t),
\label{eq:m0}
\end{equation}
which has the form of Eq.~\eqref{eq:unified} with $a_\theta=0$, $y=M_2/6$, $\Omega_\alpha^2$ replaced by
$v_{th}^2$, and $m$ replaced by $\Delta\varphi_s^{(0)}/v_{th}^2$. The long-time
self-diffusion coefficient is
\begin{equation}
D_s=\lim_{t\to\infty}\frac{M_2(t)}{6t}
=\frac{v_{th}^2}{\int_0^\infty\Delta\varphi_s^{(0)}(t)\,\dd t}.
\label{eq:ds}
\end{equation}
Equations~\eqref{eq:msd}--\eqref{eq:ds} are the same in both theories. The theories enter
only through $f_c$ and $f_s$ in $\Delta\varphi_s^{(0)}$, whose time evolutions differ, so that
$M_2(t)$ and $D_s$ also differ between ideal MCT and ideal TMCT.

\section{Critical point and the $\beta$-scaling equation}
\label{sec:beta}

\subsection{Long-time limit}

Let $f_q^\infty=\lim_{t\to\infty}f_c(q,t)$ denote the nonergodicity parameters and
$\bm f^\infty=(f_{q_1}^\infty,f_{q_2}^\infty,\dots,f_{q_M}^\infty)$ the vector of their values on the
wave-number grid $q_i$ (Sec.~\ref{sec:numerics}), and write $\mathcal{F}_q[\bm f^\infty]=\sum_{k,p}V_{qkp}f_k^\infty f_p^\infty$ for the long-time limit of $m_c$, and define
$C_{qk}=\partial \mathcal{F}_q/\partial f_k^\infty$ and $D_{qkp}=\partial^2\mathcal{F}_q/\partial f_k^\infty\partial f_p^\infty$.
For $t\gg\Omega^{-1}$ the inertia terms can be dropped, and in Laplace space, with $f_q(z)$ and
$K_q(z)$ the Laplace transforms of $f_c(q,t)$ and $K_c(q,t)$,
\begin{subequations}
\label{eq:long}
\begin{align}
\frac{zf_q(z)}{1-zf_q(z)}&=z\mathcal L[\mathcal{F}_q](z)&&(\mct),\\
\frac{1}{zK_q(z)}&=z\mathcal L[\mathcal{F}_q](z)&&(\tmct).
\end{align}
\end{subequations}
For $z\to0$, $zf_q(z)\to f_q^\infty$, $zK_q(z)\to-\ln f_q^\infty$, and
$z\mathcal L[\mathcal{F}_q](z)\to\mathcal{F}_q[\bm f^\infty]$. The nonergodicity parameters therefore
satisfy
\begin{subequations}
\label{eq:nep}
\begin{align}
f_q^\infty&=\frac{\mathcal{F}_q[\bm f^\infty]}{1+\mathcal{F}_q[\bm f^\infty]}&&(\mct),\\
f_q^\infty&=\exp\!\big(-1/\mathcal{F}_q[\bm f^\infty]\big)&&(\tmct).
\end{align}
\end{subequations}
The critical point $\varphi_c$ is where the nontrivial solution of Eq.~\eqref{eq:nep}
disappears in a fold bifurcation. We denote the nonergodicity parameters at $\varphi_c$ by
$f_q^{\infty,c}$ and the critical cumulants by $K_q^c=-\ln f_q^{\infty,c}$. The stability matrix $A$ is the Jacobian of
the fixed-point map~\eqref{eq:nep}, written in terms of $f_q^\infty$ for ideal MCT and of
$K_q^\infty=-\ln f_q^\infty$ for ideal TMCT. At $\varphi_c$ its largest eigenvalue is 1, with
\begin{subequations}
\label{eq:stab}
\begin{align}
A^\mct_{qk}&=(1-f_q^{\infty,c})^2C_{qk},\\
A^\tmct_{qk}&=(K_q^c)^2C_{qk}f_k^{\infty,c}.\label{eq:stabT}
\end{align}
\end{subequations}
Differentiating Eq.~\eqref{eq:nep} for ideal TMCT with respect to $f_k^\infty$ instead gives
$(K_q^c)^2f_q^{\infty,c}C_{qk}$, which is related to Eq.~\eqref{eq:stabT} by a similarity
transformation with the diagonal matrix of $f_q^{\infty,c}$ and has the same eigenvalues.
Since the elements of $A$ are positive, the Perron--Frobenius theorem ensures that its
largest eigenvalue is nondegenerate, with positive right and left eigenvectors $e_q$ and $\hat e_q$, normalized as $\sum_q\hat e_qe_q=1$.

\subsection{$\beta$-scaling equation of ideal TMCT}
\label{sec:betaeq}

Near $\varphi_c$ we set $K_q(t)=K_q^c+\kappa_q(t)$. The kernel is a function of $f$,
whereas the unknown is $\kappa$. Since $f=e^{-K}$, to second order,
\begin{equation}
f_k(t)-f_k^{\infty,c}=-f_k^{\infty,c}\kappa_k(t)+\tfrac12 f_k^{\infty,c}\kappa_k(t)^2+\cdots .
\label{eq:expmap}
\end{equation}
The second term is of the same order as the other quadratic terms and must be kept.
We insert Eq.~\eqref{eq:expmap} into Eq.~\eqref{eq:long} expanded around the critical point.
To first order, the terms linear in $\kappa$ combine into $(1-A)\kappa$, which vanishes only
for $\kappa_q\propto e_q$; we therefore write $\kappa_q(t)=e_qG(t)+O(G^2)$. At second order,
the correction of order $G^2$ obeys an equation with the singular operator $1-A$. Its
solvability condition, obtained by projecting onto $\hat e_q$, gives the standard
$\beta$-scaling equation
\begin{equation}
\tilde\sigma+\lambda\,z\mathcal L[G(t)^2](z)-[zG(z)]^2=0
\label{eq:betaeq}
\end{equation}
where $\tilde\sigma\propto\varphi-\varphi_c$ is the separation parameter and
\begin{equation}
\lambda_\tmct=\frac{\sum_q\hat e_q(K_q^c)^2\,\Pi_q}
{\sum_q\hat e_q\,e_q^2/K_q^c},
\label{eq:lamtmct}
\end{equation}
\begin{equation}
\Pi_q=\tfrac12\sum_kC_{qk}f_k^{\infty,c}e_k^2
+\tfrac12\sum_{k,p}D_{qkp}f_k^{\infty,c}e_kf_p^{\infty,c}e_p .
\label{eq:Pi}
\end{equation}
The derivation and the expression for $\tilde\sigma$ are given in Appendix~\ref{app:beta}.
Second order is sufficient because $G$ is of order $|\tilde\sigma|^{1/2}$ near $\varphi_c$.
The terms linear in $\kappa$ drop out on projection, so that the leading terms, $\tilde\sigma$
and those quadratic in $G$, are all of order $|\tilde\sigma|$. Cubic terms are of order
$|\tilde\sigma|^{3/2}$ and give only corrections to the asymptotic laws \cite{Gotze1991}. The
corresponding ideal MCT result is
\begin{equation}
\lambda_\mct=\frac{\tfrac12\sum_q\hat e_q(1-f_q^{\infty,c})^2\sum_{k,p}D_{qkp}e_ke_p}
{\sum_q\hat e_q\,e_q^2/(1-f_q^{\infty,c})},
\label{eq:lammct}
\end{equation}
which is equivalent to the standard expression \cite{Gotze1991,Franosch1997}.

Because Eq.~\eqref{eq:betaeq} has the same form as in ideal MCT, the asymptotic laws follow
as there \cite{Gotze1991}. To leading order, $f_c(q,t)-f_q^{\infty,c}$ is proportional to
$h_qG(t)$, with the critical amplitude $h_q=e_q$ in ideal MCT and $h_q=f_q^{\infty,c}e_q$ in
ideal TMCT [Eq.~\eqref{eq:expmap}]. At $\varphi_c$, $G$ decays as $t^{-a}$ (critical decay);
below $\varphi_c$, the decay from the plateau follows the von Schweidler law, $G\propto-t^{b}$.
The exponents are fixed by $\lambda$ through
\begin{equation}
\frac{\Gamma(1-a)^2}{\Gamma(1-2a)}=\lambda=\frac{\Gamma(1+b)^2}{\Gamma(1+2b)},
\label{eq:ab}
\end{equation}
and the $\alpha$-relaxation time diverges as $\tau\propto|\tilde\sigma|^{-\gamma}$ with
$\gamma=1/(2a)+1/(2b)$. As in ideal MCT, we take $D_s\propto1/\tau$, so that
$D_s\propto(1-\varphi/\varphi_c)^\gamma$ near $\varphi_c$.

\subsection{Structure of the additional term}
\label{sec:struct}

The first term of $\Pi_q$ has no counterpart in ideal MCT. By
Eq.~\eqref{eq:stabT}, $\sum_q\hat e_q(K_q^c)^2C_{qk}f_k^{\infty,c}=\sum_q\hat e_qA^\tmct_{qk}$, and
this equals $\hat e_k$ because $\hat e_q$ is the left eigenvector of $A^\tmct$ with eigenvalue 1.
The first term therefore simplifies, and
\begin{equation}
\lambda_\tmct=\tfrac12\langle K^c\rangle_w+\lambda_D,
\label{eq:lamsplit}
\end{equation}
with
\begin{equation}
\langle K^c\rangle_w=\frac{\sum_q w_qK_q^c}{\sum_qw_q},\qquad w_q=\frac{\hat e_qe_q^2}{K_q^c},
\end{equation}
where $\lambda_D$ is the $D$ term of Eq.~\eqref{eq:lamtmct}. The additional contribution is
half a weighted average of the critical cumulant. Because $e_q$, $\hat e_q$, and $K_q^c$ are
positive, the weights $w_q$ are positive, so that $\langle K^c\rangle_w$ is a proper average
and the additional term always increases $\lambda$: $\lambda_\tmct>\lambda_D$. It vanishes only if $K_q^c\to0$ ($f_q^{\infty,c}\to1$)
for all $q$. In the same limit, $K^c\simeq1-f^{\infty,c}$ and $\lambda_D$ reduces to
Eq.~\eqref{eq:lammct}, so that $\lambda_\tmct\to\lambda_\mct$. The same result follows if
the kernel is written in terms of the cumulants,
$\mathcal{F}_q=\sum_{k,p}V_{qkp}f_k^{\infty,c}f_p^{\infty,c}e^{-(\kappa_k+\kappa_p)}$, and the exponential is expanded:
the additional term stems from the diagonal terms $\kappa_k^2$. An expression for $\lambda$
of ideal TMCT has been given in Ref.~\cite{Tokuyama2026} by expanding $f$ to first order in
$\kappa$. The additional term in Eq.~\eqref{eq:lamsplit} arises from the second-order term in
Eq.~\eqref{eq:expmap}, which is of the same order as the other quadratic terms.

\subsection{Schematic models}

Schematic models neglect the wave-vector dependence and describe a single correlator $f(t)$
with a memory kernel $\mathcal{F}[f]$. If $S(q)$ is approximated by a $\delta$ function at its
main peak, only one wave number remains, and the kernel of Eq.~\eqref{eq:mc} reduces to the
quadratic form $\mathcal{F}=vf^2$ \cite{Bengtzelius1984}. Kernels with other powers, such as
$\mathcal{F}=vf^n$ or $\mathcal{F}=v_1f+v_3f^3$, are not derived from a structure factor; they are
model kernels used to study how $\lambda$ depends on the form of $\mathcal{F}$ \cite{GotzeBook}.

For a single correlator, the stability matrix is a number and $\hat e=e=1$. Equations
\eqref{eq:lamtmct}--\eqref{eq:lammct} then reduce to
\begin{subequations}
\label{eq:lamschem}
\begin{align}
\lambda_\mct&=\tfrac12(1-f^{\infty,c})^3\mathcal{F}''&&(\mct),\\
\lambda_\tmct&=\tfrac12(K^c)^3\big[f^{\infty,c}\mathcal{F}'+(f^{\infty,c})^2\mathcal{F}''\big]&&(\tmct),
\end{align}
\end{subequations}
with the derivatives taken at $f^{\infty,c}$. The critical point is given by Eq.~\eqref{eq:nep}
together with $A=1$, i.e., $(1-f^{\infty,c})^2\mathcal{F}'=1$ in ideal MCT and
$(K^c)^2f^{\infty,c}\mathcal{F}'=1$ in ideal TMCT. The latter condition turns the first term in the
bracket into $\tfrac12K^c$, the additional term of Eq.~\eqref{eq:lamsplit}.

For a single power, $\mathcal{F}=vf^n$ with $n\ge2$, Eq.~\eqref{eq:nep} gives
$vf^{n-1}=1/(1-f^{\infty,c})$ in ideal MCT and $vf^n=1/K^c$ in ideal TMCT, so that the
condition $A=1$ becomes $n(1-f^{\infty,c})=1$ and $nK^c=1$, respectively. Inserting these into
Eq.~\eqref{eq:lamschem} gives $\lambda=1/2$ in both theories, for every $n$. In ideal TMCT this
value is the sum $\tfrac12K^c+\lambda_D=1/(2n)+(n-1)/(2n)$: the additional term is present, but
$\lambda_D$ is smaller than $\lambda_\mct$ by the same amount. For a single power, the critical
conditions thus depend on $n$ only, and the plateau differs between the theories,
$f^{\infty,c}=1-1/n$ in ideal MCT and $e^{-1/n}$ in ideal TMCT; inserting these values gives the
same $\lambda=1/2$ in both.

A difference in $\lambda$ appears as soon as the kernel contains more than one power of $f$.
For $\mathcal{F}=v_1f+v_3f^3$, the same steps give $\lambda_\tmct=K^c(2-\tfrac32K^c)$, the result
of G\"otze and Schilling \cite{GotzeSchilling2015}, and $\lambda_\mct=\tfrac32(1-f^{\infty,c})$.
For small $K^c\simeq1-f^{\infty,c}$, $\lambda_\tmct-\lambda_\mct\simeq\tfrac12K^c$, which is the
additional term. In single-power models, the additional term is
compensated by $\lambda_D$ and does not show up in $\lambda$, so that these models cannot
distinguish the exponents of the two theories. The difference becomes visible in models with several powers and, as
shown in Sec.~\ref{sec:results}, in the wave-vector-dependent theory.

\section{Numerical method}
\label{sec:numerics}

The numerical code described in this section was written and run with the assistance of an
AI tool; its use and the checks of its output are described in the AI Use Disclosure.

\subsection{Static input and wave-vector integrals}

The Fourier transform $c(q)$ of the PY direct correlation function is known in closed form
\cite{HansenMcDonald}. We evaluate it instead by 400-point Gauss--Legendre quadrature of the
analytic $c(r)$, which agrees with the closed form to $10^{-12}$ and avoids its loss of
precision by cancellation at small $q$, and obtain $S(q)$ from the Ornstein--Zernike relation. Wave numbers are discretized as $q_i=(i+1/2)\Delta q$
($i=0,\dots,M-1$) with $\Delta q\sigma=0.2$, so that $q_c=M\Delta q$. In the memory kernels
\eqref{eq:mc} and \eqref{eq:ms}, the integration over $\bm k$ is transformed to the bipolar
coordinates $k=|\bm k|$ and $p=|\bm q-\bm k|$ \cite{Franosch1997},
\begin{equation}
\int\dd^3k\,F(k,p)=\frac{2\pi}{q}\int_0^{q_c}\dd k\,k\int_{|q-k|}^{q+k}\dd p\,p\,F(k,p),
\label{eq:bipolar}
\end{equation}
where $p$ is also restricted to $p<q_c$, and the double integral is replaced by a sum over the
grid points. The kernel then takes the
form $m_c(q_i,t)=\sum_{k,p}V_{q_ikp}f_c(k,t)f_c(p,t)$, with a vertex $V_{qkp}$ that is computed
once for each $\varphi$; $m_s$ and $\Delta\varphi_s^{(0)}$ are discretized in the same way.

\subsection{Nonergodicity parameters and critical point}

At a given $\varphi$, Eq.~\eqref{eq:nep} is solved by iteration starting from
$\bm f^\infty=\bm1$. The iteration is stopped when $\max_q|\Delta f_q^\infty|<10^{-13}$ (glass)
or when all $f_q^\infty$ have fallen below $10^{-4}$ (liquid). Near $\varphi_c$ the number of
iterations grows as $|\varphi-\varphi_c|^{-1/2}$ on both sides, so that we do not limit it:
an iteration stopped early misclassifies slowly collapsing liquid states.

The critical point is first bracketed by bisection in $\varphi$ to $10^{-6}$. We then compute the glass-side fixed
points at nine volume fractions between $2\times10^{-6}$ and $2\times10^{-4}$ above the
bracket. At each of them we compute $C_{qk}=\sum_p(V_{qkp}+V_{qpk})f_p^\infty$ and
$D_{qkp}=V_{qkp}+V_{qpk}$, construct the stability matrix from Eq.~\eqref{eq:stab} with
$f_q^\infty$ and $K_q^\infty$ in place of their critical values, and obtain its largest
eigenvalue $E$ and the corresponding right and left eigenvectors $e_q$ and $\hat e_q$,
normalized as $\sum_q\hat e_qe_q=1$; the eigenvectors are obtained by full diagonalization
of $A$ and its transpose, whose dimension is at most 300. Since $E$ satisfies $1-E\propto(\varphi-\varphi_c)^{1/2}$ near a
fold bifurcation, we obtain $\varphi_c$ from a fit to $1-E=c_1s+c_2s^2+c_3s^3$ with
$s=(\varphi-\varphi_c)^{1/2}$. Inserting the same quantities into
Eqs.~\eqref{eq:lamtmct}--\eqref{eq:lammct} gives $\lambda$ at each fixed point, and its value
at $\varphi_c$ is obtained by extrapolating a quadratic fit in $s$ to $s=0$. The critical fixed
point itself is not used, because the iteration does not converge in the limit
$\varphi\to\varphi_c^+$. The two contributions in Eq.~\eqref{eq:lamsplit}, $\tfrac12\langle K^c\rangle_w$ and
$\lambda_D$, are evaluated in the same way from the two terms of $\Pi_q$ at three volume
fractions $10^{-7}$--$10^{-6}$ above $\varphi_c$ and extrapolated to $s=0$ separately; their
sum agrees with $\lambda$ to the quoted digits. In
ideal TMCT, modes with $f_q^\infty<10^{-100}$ are omitted from $A$ and $\Pi_q$ to avoid
numerical underflow.

As an independent check, $\lambda$ is also obtained from the glass-side fixed points without
using $\Pi_q$. On the glass side, $G(t)$ tends to a constant $g$, and Eq.~\eqref{eq:betaeq}
with $z\mathcal L[G^2]\to g^2$ and $zG\to g$ gives $g=[\tilde\sigma/(1-\lambda)]^{1/2}$. We
compute the exact fixed points at seven volume fractions with
$10^{-6}\le\varphi-\varphi_c\le6.4\times10^{-5}$ and form their deviations from the critical
fixed point, $\delta_q=f_q^\infty-f_q^{\infty,c}$ (ideal MCT) or $\delta_q=K_q^\infty-K_q^c$
(ideal TMCT). Since $\delta_q=e_qg+O(g^2)$ and $\sum_q\hat e_qe_q=1$, the amplitude follows as
$g=\sum_q\hat e_q\delta_q$, and $\tilde\sigma$ follows from Eq.~\eqref{eq:sigt}. A fit of
$g=\alpha\tilde\sigma^{1/2}+\beta\tilde\sigma+c$, where $\beta$ accounts for the next-order
correction and $c$ for the error of the critical fixed point, gives $\lambda=1-1/\alpha^2$.

\subsection{Time evolution}

Equation~\eqref{eq:unified} for $f_c$ and $f_s$ at all $q_i$ is integrated together with
Eq.~\eqref{eq:msd} by the decimation algorithm of Fuchs \textit{et al.} \cite{Fuchs1991}. The
time axis is covered by blocks of $N$ points with a constant step $h$. The first half of the
first block uses the short-time solution $y=\Omega_\alpha^2t^2/2$. Within a block, the memory
integral is discretized with the averages of the kernel over each step \cite{Fuchs1991}, the
second derivative with a second-order backward difference, and the resulting implicit
equation for $y(t_i)$ is solved by iteration to a relative accuracy of $10^{-12}$. After each
block, every second point is kept, $h$ is doubled, and the second half of the next block is
computed. We use $N=256$ points per block, an initial step $h_0=10^{-7}t_0$, and final times up
to $10^{14}t_0$; for the direct tests of the $\beta$-relaxation exponents
(Sec.~\ref{sec:results}) we use $N=1024$. The long-time self-diffusion coefficient is
obtained from both expressions in Eq.~\eqref{eq:ds}.

\subsection{Numerical accuracy}

We examined how the two discretization parameters, the wave-number grid spacing $\Delta q$
and the number $N$ of time points per block, affect the results.

\textit{Grid spacing.} The static results are converged for $\Delta q\sigma\le0.2$
(Table~\ref{tab:dq}): halving $\Delta q$ to $0.1\sigma^{-1}$ changes $\varphi_c$ by less than
$5\times10^{-5}$ and $\lambda$ by less than $2\times10^{-4}$, and the position of $q^*$ does not
depend on $\Delta q$. The coarser grid $\Delta q\sigma=0.4$ underestimates $\varphi_c$ of ideal
TMCT by $7\times10^{-4}$ but still gives $\lambda$ within $0.003$. For $D_s$ of ideal TMCT,
$\Delta q\sigma=0.1$ and 0.2 agree to $0.2\%$ for $10^{-2}\le1-\varphi/\varphi_c\le0.15$, whereas
$\Delta q\sigma=0.4$ overestimates $D_s$ by $2\%$ at $1-\varphi/\varphi_c=0.15$ and by $27\%$ at
$10^{-2}$. The lowest grid point is affected by the discretization and is omitted from the
figures; removing it changes neither $\varphi_c$ nor $\lambda$ within the quoted digits.

\textit{Time discretization.} Increasing $N$ from 256 to 1024 changes $D_s$ by less than 1\%
for $\varphi\ge0.2$, and the two expressions in Eq.~\eqref{eq:ds} agree to four digits. The only
quantity sensitive to $N$ is the local exponent of the critical decay at late times, which
drifts downward. Each doubling of $N$ from 256 to 2048 postpones this drift by about one decade
in time; for ideal MCT with $N=2048$ the local exponent is $0.31$--$0.32$ throughout the window
$10^2<t/t_0<10^5$ used in Sec.~\ref{sec:results}.

\textit{A remaining limitation.} For ideal MCT at $\varphi\lesssim0.1$, $f_s(q,t)$ oscillates
almost without damping when $\gamma_s=0$, and $D_s$ does not converge with $N$. We therefore
show ideal MCT results only for $\varphi\ge0.2$. Ideal TMCT does not have this problem.

\begin{table}
\caption{Convergence of $\varphi_c$ and $\lambda$ with respect to the grid spacing $\Delta q$ at
$q_c=q^*$.}
\label{tab:dq}
\begin{ruledtabular}
\begin{tabular}{ccccc}
 & \multicolumn{2}{c}{Ideal MCT} & \multicolumn{2}{c}{Ideal TMCT}\\
$\Delta q\sigma$ & $\varphi_c$ & $\lambda$ & $\varphi_c$ & $\lambda$\\
\hline
0.4 & 0.51654 & 0.7074 & 0.58174 & 0.8575\\
0.2 & 0.51650 & 0.7069 & 0.58244 & 0.8550\\
0.1 & 0.51646 & 0.7069 & 0.58241 & 0.8551\\
\end{tabular}
\end{ruledtabular}
\end{table}

\section{Results}
\label{sec:results}

\subsection{Cutoff dependence and the intrinsic cutoff of ideal TMCT}

\begin{figure}
\includegraphics[width=\columnwidth]{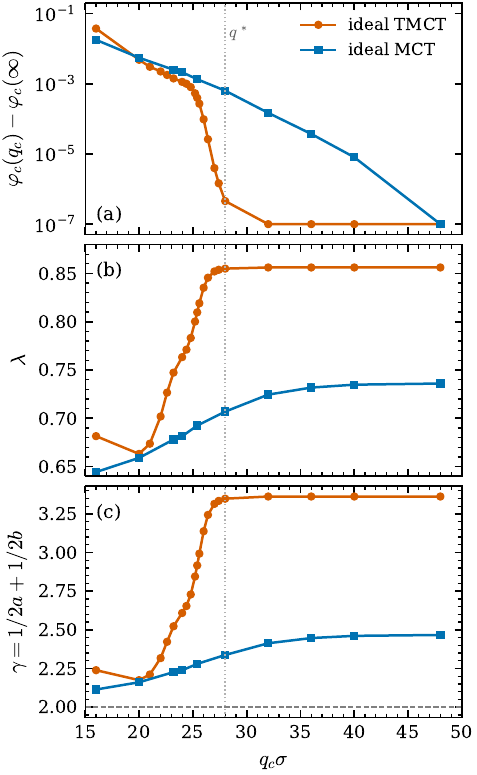}
\caption{Dependence on the wave-vector cutoff $q_c$ (grid spacing $\Delta q\sigma=0.2$).
(a) Deviation of the critical volume fraction from its large-$q_c$ limit,
$\varphi_c(\infty)=0.582435$ (ideal TMCT) and 0.515867 (ideal MCT, the value at the
largest cutoff $q_c\sigma=48$, used as a reference); values below $10^{-7}$ are
plotted at $10^{-7}$. (b) Exponent parameter. (c) Exponent $\gamma=1/(2a)+1/(2b)$ of
$D_s$; the dashed line marks $\gamma=2$. Circles: ideal TMCT; squares: ideal MCT. The dotted line
marks $q^*\sigma=28$. This and all following figures were produced from the numerical data
of this work with plotting scripts written by an AI tool (see AI Use Disclosure).}
\label{fig:cutoff}
\end{figure}

Figure~\ref{fig:cutoff} shows $\varphi_c$, $\lambda$, and $\gamma$ as functions of $q_c$.
In ideal MCT all three converge slowly: $\varphi_c$ is constant to four digits
only for $q_c\sigma\gtrsim36$, and $\lambda$ reaches $0.736$ at $q_c\sigma=48$. Since
$\varphi_c$ of ideal MCT still changes by $8\times10^{-6}$ between $q_c\sigma=40$ and 48, its
large-$q_c$ limit is not reached in the range examined; $\varphi_c(\infty)$ for ideal MCT in
Fig.~\ref{fig:cutoff}(a) is the value at the largest cutoff, $q_c\sigma=48$, which serves only
as a reference. In ideal TMCT, $\varphi_c$ decreases smoothly from 0.5873 at $q_c\sigma=20$ and then stops changing:
$\varphi_c(q_c)-\varphi_c(\infty)$ falls below $10^{-6}$ at $q_c\sigma=28$, and
$\varphi_c=0.582435$ for all $q_c\sigma$ from 28 to 48. The same holds for $\lambda$, which
saturates at 0.855--0.856.

\begin{figure}
\includegraphics[width=\columnwidth]{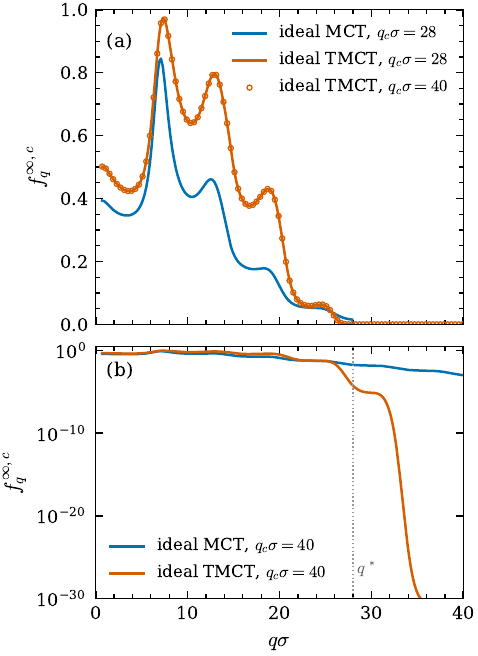}
\caption{Nonergodicity parameters at $\varphi_c$. (a) Linear scale, for $q_c\sigma=28$
(lines) and, for ideal TMCT, $q_c\sigma=40$ (symbols, every second point). (b) Logarithmic
scale, for $q_c\sigma=40$. The ideal TMCT values are exponentially small for $q\sigma\gtrsim28$
(dotted line), whereas the ideal MCT values decay slowly.}
\label{fig:fq}
\end{figure}

The origin of this intrinsic cutoff is visible in the nonergodicity parameter
(Fig.~\ref{fig:fq}). Because $f_q^{\infty,c}=\exp(-1/\mathcal{F}_q)$, it vanishes faster than any power of
$\mathcal{F}_q$ once $\mathcal{F}_q$ becomes small at large $q$. Computed with a large cutoff, $f_q^{\infty,c}$ drops from
$1.3\times10^{-2}$ at $q\sigma=26.3$ to $2.8\times10^{-5}$ at $q\sigma=28.3$ and below
$10^{-29}$ for $q\sigma>35$. Density fluctuations on length scales shorter than $\sim1/q^*$
relax completely even at $\varphi_c$; they do not couple to the frozen modes and drop out of
Eq.~\eqref{eq:nep}. We therefore identify $q^*\sigma\simeq28$ as the intrinsic cutoff of
ideal TMCT.

This cutoff corresponds to the maximum critical cutoff $1/b_c$ of Tokuyama and Narumi, who
restricted the wave numbers to $k\le1/b$ (so that $q_c=1/b$) and found $\sigma/b_c=27$ and
$\varphi_c\simeq0.5817$ for all $b\le b_c$ \cite{TokuyamaNarumi2019}. This value was first
reported in Ref.~\cite{KimuraTokuyama2016}, together with $\varphi_c\simeq0.5856$ at
$q_c\sigma=20$ and the ideal MCT values 0.5159 and 0.5214 at $q_c\sigma=40$ and 20. Our calculation
with $\Delta q\sigma=0.4$ reproduces all four values to the quoted digits, whereas the
converged grid gives $\varphi_c=0.5824$ (Sec.~\ref{sec:numerics}). Tokuyama and Narumi
attributed the existence of $b_c$ to $S(q)$ approaching unity at moderate $q$. Equation~\eqref{eq:nep} shows
why this leads to a sharp cutoff in ideal TMCT but not in ideal MCT: for small $\mathcal{F}_q$, $f_q^{\infty,c}\simeq \mathcal{F}_q$
in ideal MCT, whereas $f_q^{\infty,c}=e^{-1/\mathcal{F}_q}$ is exponentially small in ideal TMCT.
Our results add that $\lambda$, and hence all asymptotic exponents, become independent of
the cutoff at the same point.

Below $q^*$, the exponents are much more sensitive to the cutoff than the critical point
(Table~\ref{tab:crit}). Between $q_c\sigma=20$ and 28, $\varphi_c$ changes only from 0.5873
to 0.5824, while $\lambda$ changes from 0.663 to 0.855 and $\gamma$ from 2.17 to 3.35. Most
of this change occurs between $q_c\sigma\simeq23$ and 27, where $\varphi_c$ already lies
within $1.5\times10^{-3}$ of its limit. Since $\varphi_c$ decreases monotonically with $q_c$,
the large-cutoff value 0.5824 is the smallest $\varphi_c$ that ideal TMCT gives in the
range examined ($q_c\sigma\ge16$). It is close to the most recent molecular-dynamics
estimate for monodisperse hard spheres, 0.5820 \cite{TokuyamaNarumi2019}; since $\varphi_c$
of ideal TMCT does not fall below 0.5824, no cutoff reproduces this value exactly. To
illustrate the sensitivity below $q^*$, we also show results for $q_c\sigma=25.4$, where
$\varphi_c=0.5828$ exceeds its large-cutoff value by only $4\times10^{-4}$. This small shift is
accompanied by $\lambda=0.810$ and $\gamma=2.92$, i.e., a change of $\gamma$ by 0.4. In
the following we use $q_c=q^*=28\sigma^{-1}$, where the results of ideal TMCT are cutoff
independent, and show results for $q_c\sigma=25.4$ for comparison.

The decomposition~\eqref{eq:lamsplit} shows where this sensitivity comes from. At
$q_c=q^*$ we find $\tfrac12\langle K^c\rangle_w=0.595$ and $\lambda_D=0.260$; at
$q_c\sigma=25.4$, $\tfrac12\langle K^c\rangle_w=0.551$ and $\lambda_D=0.258$. The $D$ term,
which corresponds to the ideal MCT expression, hardly changes. The change of $\lambda_\tmct$ comes
almost entirely from the additional term: as the cutoff is lowered, the modes with large
$K_q^c$ near $q^*$ are removed and $\langle K^c\rangle_w$ decreases.

\begin{table}
\caption{Critical volume fraction, exponent parameter, exponents $a$ and $b$, and
$\gamma=1/(2a)+1/(2b)$. For ideal TMCT the values for $q_c\ge q^*$ are those of $q_c\sigma=28$;
$q_c\sigma=40$ and 48 give $\varphi_c=0.5824347$ and $\lambda=0.8562$. Grid spacing
$\Delta q\sigma=0.2$; the digits of $\varphi_c$ beyond the fourth depend on $\Delta q$
(Sec.~\ref{sec:numerics}).}
\label{tab:crit}
\begin{ruledtabular}
\begin{tabular}{llccccc}
Theory & $q_c\sigma$ & $\varphi_c$ & $\lambda$ & $a$ & $b$ & $\gamma$\\
\hline
Ideal MCT  & 20   & 0.5213790 & 0.6590 & 0.343 & 0.710 & 2.16\\
Ideal MCT  & 25.4 & 0.5172101 & 0.6924 & 0.330 & 0.653 & 2.28\\
Ideal MCT  & 28   & 0.5165025 & 0.7069 & 0.324 & 0.629 & 2.34\\
Ideal MCT  & 48   & 0.5158671 & 0.7360 & 0.311 & 0.581 & 2.47\\
Ideal TMCT & 20   & 0.5872533 & 0.6631 & 0.342 & 0.703 & 2.17\\
Ideal TMCT & 25.4 & 0.5828307 & 0.8097 & 0.273 & 0.460 & 2.92\\
Ideal TMCT & 28   & 0.5824352 & 0.8550 & 0.245 & 0.383 & 3.35\\
\end{tabular}
\end{ruledtabular}
\end{table}

\subsection{Exponent parameter}
\label{sec:lam}

\begin{figure*}
\includegraphics[width=\textwidth]{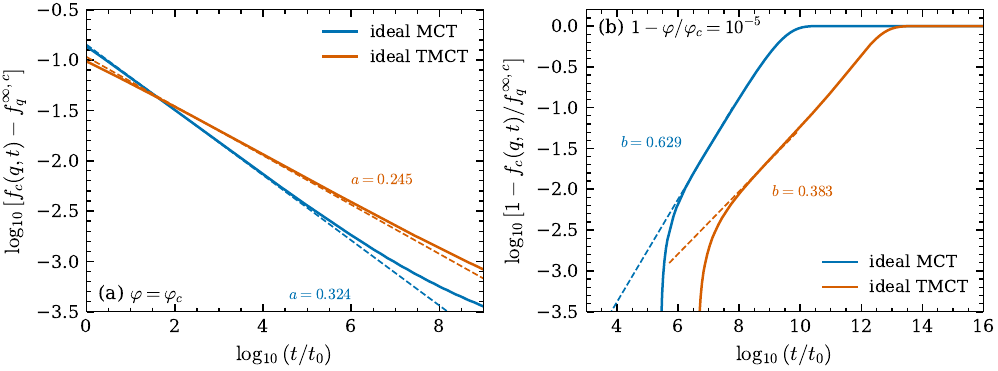}
\caption{Direct test of the $\beta$-relaxation exponents at $q_c=q^*$ and $q\sigma=10.5$.
(a) Critical decay at $\varphi=\varphi_c$. (b) von Schweidler law at
$1-\varphi/\varphi_c=10^{-5}$. Dashed lines: power laws with the exponents predicted from
$\lambda$ (Table~\ref{tab:crit}).}
\label{fig:beta}
\end{figure*}

Table~\ref{tab:lam} summarizes four independent determinations of $\lambda$ at $q^*$.
\begin{enumerate}
\item \textit{Stability matrix:} Eqs.~\eqref{eq:lamtmct} and \eqref{eq:lammct}, evaluated as
described in Sec.~\ref{sec:numerics}. For ideal TMCT,
the additional term $\tfrac12\langle K^c\rangle_w=0.595$ of Eq.~\eqref{eq:lamsplit} is
larger than $\lambda_D=0.260$; without it, $\lambda$ would be 0.26.
\item \textit{Static fixed-point analysis:} the relation $g=[\tilde\sigma/(1-\lambda)]^{1/2}$
between the amplitude $g$ of the glass-side fixed points and $\tilde\sigma$, which gives
$\lambda$ without using $\Pi_q$ (Sec.~\ref{sec:numerics}).
\item \textit{Critical decay} at $\varphi=\varphi_c$ [Fig.~\ref{fig:beta}(a)].
\item \textit{von Schweidler law} at $1-\varphi/\varphi_c=10^{-5}$ [Fig.~\ref{fig:beta}(b)].
\end{enumerate}
The static analysis reproduces the stability-matrix values to $3\times10^{-4}$. For the
dynamical tests we use $q\sigma=4.1$ and 10.5, on either side of the main peak of $S(q)$,
where $f_q^{\infty,c}=0.43$ and 0.64 in ideal TMCT (0.35 and 0.41 in ideal MCT) is far from
unity. The critical amplitude is then large, and the leading asymptotic laws are not masked by
corrections (see below). The dynamical exponents agree with the predictions within the accuracy expected from
preasymptotic corrections. For ideal TMCT the measured local exponents are $a=0.24$ and
$b=0.40$--$0.45$ (at $q\sigma=4.1$ and 10.5), compared with the predicted $a=0.245$ and
$b=0.383$. The local exponent of ideal TMCT at $q\sigma=10.5$ increases from 0.23 at
$t/t_0=10$ to 0.241 at $10^5$, approaching the predicted value 0.245 from below. For ideal MCT
they are $a=0.29$--$0.32$ and $b=0.61$--$0.66$, compared with 0.324 and
0.629. For each theory, the exponents predicted from the $\lambda$ of the other theory
($a=0.324$ and $b=0.629$ for ideal TMCT; $a=0.245$ and $b=0.383$ for ideal MCT) lie outside the
measured ranges. For ideal TMCT, the same holds for the exponents $a\simeq0.313$ and
$b\simeq0.587$ that correspond to the earlier estimate $\lambda\simeq0.733$ \cite{Tokuyama2026}.
The downward drift of the local exponent of the critical decay at late times
[Fig.~\ref{fig:beta}(a)] is a discretization error (Sec.~\ref{sec:numerics}).

\begin{table}
\caption{Determinations of $\lambda$ and of the exponents at $q_c=q^*=28\sigma^{-1}$, and
at $q\sigma=4.1$ and 10.5 for the dynamical tests. The measured local exponents, $-\dd\ln|f_c-f_q^{\infty,c}|/\dd\ln t$ (critical decay) and
$\dd\ln(1-f_c/f_q^{\infty,c})/\dd\ln t$ (von Schweidler law), refer to
$10^2<t/t_0<10^5$ (critical decay) and $0.01<1-f/f^{\infty,c}<0.1$ (von Schweidler law), with
$N=1024$.}
\label{tab:lam}
\begin{ruledtabular}
\begin{tabular}{lcc}
 & Ideal MCT & Ideal TMCT\\
\hline
$\lambda$ (stability matrix) & 0.7069 & 0.8550\\
$\lambda$ (static analysis) & 0.7070 & 0.8553\\
$\tfrac12\langle K^c\rangle_w$ / $\lambda_D$ & --- & 0.595 / 0.260\\
$a$ predicted / measured & 0.324 / 0.29--0.32 & 0.245 / 0.24\\
$b$ predicted / measured & 0.629 / 0.61--0.66 & 0.383 / 0.40--0.45\\
\end{tabular}
\end{ruledtabular}
\end{table}

Near the main peak of $S(q)$, ideal TMCT gives $f_q^{\infty,c}\simeq0.97$. Since $e_q$ contains the
factor $(K_q^c)^2$ by Eq.~\eqref{eq:stabT}, the critical amplitude $h_q=f_q^{\infty,c}e_q$ is then
very small, and the leading asymptotic laws are masked
by corrections. At $q\sigma=7.3$ and $1-\varphi/\varphi_c=10^{-5}$, the local von Schweidler
exponent is $0.41$--$0.44$ only for $1-f/f^{\infty,c}<0.01$; for larger deviations from the plateau it
is $0.54$--$0.63$. At $\varphi=0.58$ ($1-\varphi/\varphi_c=4.2\times10^{-3}$) it is
$0.57$--$0.67$ at this wave vector, and a von Schweidler fit with $b=0.583$, the value that
corresponds to $\lambda=0.735$, describes the decay better than one with the true exponent
$b=0.383$. Fits near the main peak at accessible distances from $\varphi_c$ thus yield an
apparent $\lambda$ close to that of ideal MCT. The earlier estimate $\lambda\simeq0.733$
\cite{Tokuyama2026} was obtained from exponents $a\simeq0.313$ and $b\simeq0.587$ at a wave
vector with $f^{\infty,c}\simeq0.973$, and is consistent with this apparent value.

\subsection{Intermediate scattering functions and mean-square displacement}

\begin{figure}
\includegraphics[width=\columnwidth]{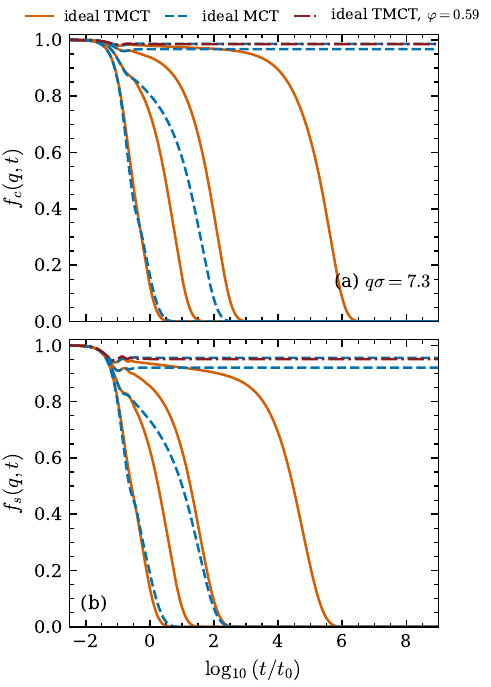}
\caption{(a) Collective and (b) self intermediate scattering functions at $q\sigma=7.3$ for
$\varphi=0.40$, 0.50, 0.55, and 0.58 (left to right) at $q_c=q^*$. Solid lines: ideal TMCT;
dashed lines: ideal MCT; dash-dotted lines: ideal TMCT in the glass state at $\varphi=0.59$, whose
plateau nearly coincides with that of ideal MCT at $\varphi=0.58$.}
\label{fig:fqt}
\end{figure}

Figure~\ref{fig:fqt} compares $f_c(q,t)$ and $f_s(q,t)$ near the main peak of $S(q)$. At
equal $\varphi$ the ideal TMCT correlators relax much faster than the ideal MCT ones. At $\varphi=0.55$
and 0.58, ideal MCT is in the glass state, whereas ideal TMCT still relaxes. At $\varphi=0.58$ the
ideal TMCT relaxation time, defined by $f_\alpha(q,\tau)=e^{-1}$, is $\tau\simeq3\times10^5t_0$ for the
collective and $5\times10^4t_0$ for the self correlator. At $\varphi=0.50$, where both
theories are ergodic, the collective $\tau$ of ideal MCT is six times that of ideal TMCT.
Above its critical point, at $\varphi=0.59$, ideal TMCT also shows a permanent plateau, with
$f_c=0.985$ and $f_s=0.951$ at $q\sigma=7.3$. These values nearly coincide with those of
ideal MCT at $\varphi=0.58$ (0.986 and 0.956), although the two states lie at very different
distances from their critical points (1.3\% and 12\% above $\varphi_c$).

\begin{figure}
\includegraphics[width=\columnwidth]{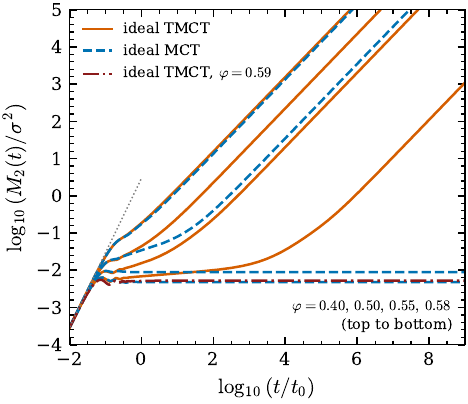}
\caption{Mean-square displacement for $\varphi=0.40$, 0.50, 0.55, and 0.58 (top to bottom)
at $q_c=q^*$. Solid lines: ideal TMCT; dashed lines: ideal MCT; dash-dotted line: ideal TMCT in
the glass state at $\varphi=0.59$.}
\label{fig:msd}
\end{figure}

Figure~\ref{fig:msd} shows $M_2(t)$. The ballistic regime
$M_2=3v_{th}^2t^2$ is common to both theories. A cage plateau develops as $\varphi$
increases; for $\varphi>\varphi_c^\mct$, ideal MCT gives a permanent plateau, whereas ideal TMCT remains
diffusive. At $\varphi=0.58$ the ideal TMCT plateau, $M_2\simeq0.008$--$0.010\sigma^2$ for
$10<t/t_0<100$, corresponds to a localization length $(M_2/6)^{1/2}\simeq0.04\sigma$.
This is the same as for ideal MCT at $\varphi=0.55$ ($0.039\sigma$). In the glass state at
$\varphi=0.59$, ideal TMCT gives a permanent plateau $M_2=0.0053\sigma^2$, close to that of ideal
MCT at $\varphi=0.58$ ($0.0048\sigma^2$).

\begin{figure}
\includegraphics[width=\columnwidth]{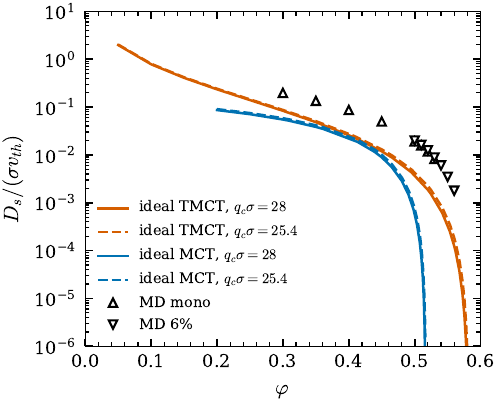}
\caption{Long-time self-diffusion coefficient versus $\varphi$ at $q_c=q^*$ (solid) and
$q_c\sigma=25.4$ (dashed). Ideal MCT is shown for $\varphi\ge0.2$ only (Sec.~\ref{sec:numerics}).
Triangles: molecular dynamics of monodisperse and 6\% polydisperse hard spheres
\cite{Tokuyama2003}, converted to units of $\sigma v_{th}$. The theories are for
$\gamma_\alpha=0$; the magnitudes are not meant to be compared with the simulations
(Sec.~\ref{sec:discussion}).}
\label{fig:ds}
\end{figure}

\begin{figure*}
\includegraphics[width=\textwidth]{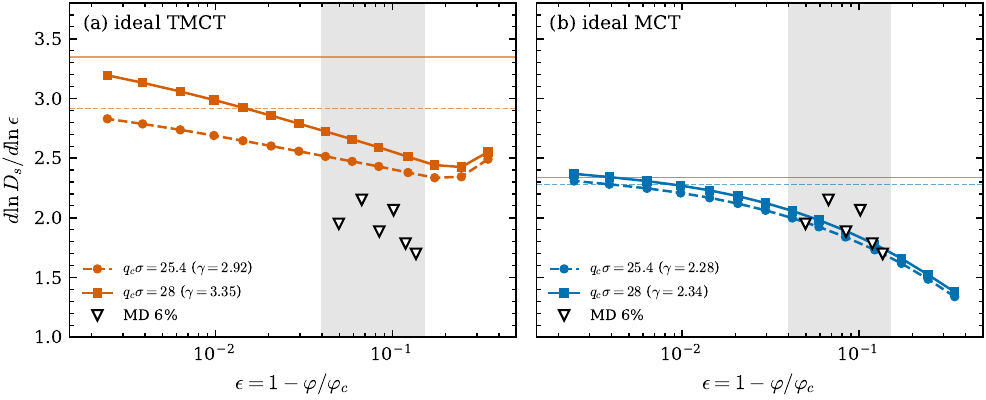}
\caption{Local exponent $\dd\ln D_s/\dd\ln\epsilon$, $\epsilon=1-\varphi/\varphi_c$, for
(a) ideal TMCT and (b) ideal MCT at $q_c\sigma=25.4$ (circles, dashed) and $q^*\sigma=28$ (squares,
solid). Horizontal lines: asymptotic values $\gamma=1/(2a)+1/(2b)$. Shaded band: range
typically used to fit simulation data. Triangles: molecular dynamics of 6\% polydisperse
hard spheres \cite{Tokuyama2003}, with $\varphi_c=0.5845$ \cite{Tokuyama2007}.}
\label{fig:lgam}
\end{figure*}

\subsection{Long-time self-diffusion: asymptotic and effective exponents}

\begin{table}
\caption{Cutoff dependence of $D_s$ above $q^*$: ratio $D_s(q_c)/D_s(q^*)$ at $\varphi=0.20$
and at $\epsilon=0.15$ and 0.04, and effective exponent $\gamma_{\rm eff}$ for
$0.04\le\epsilon\le0.15$ ($\gamma_{\rm eff}=1.91$ for ideal MCT and 2.60 for ideal TMCT at $q^*$).}
\label{tab:qcdep}
\begin{ruledtabular}
\begin{tabular}{llcccc}
 & $q_c\sigma$ & $\varphi=0.20$ & $\epsilon=0.15$ & $\epsilon=0.04$ & $\gamma_{\rm eff}$\\
\hline
Ideal MCT & 32 & 0.88 & 0.83 & 0.75 & 1.99\\
 & 40 & 0.78 & 0.65 & 0.51 & 2.09\\
 & 48 & 0.67 & 0.51 & 0.37 & 2.16\\
Ideal TMCT & 32 & 0.95 & 0.81 & 0.59 & 2.84\\
 & 40 & 0.88 & 0.62 & 0.32 & 3.10\\
 & 48 & 0.83 & 0.49 & 0.21 & 3.26\\
 & 60 & 0.77 & 0.40 & 0.14 & 3.40\\
\end{tabular}
\end{ruledtabular}
\end{table}

Figure~\ref{fig:ds} shows $D_s(\varphi)$. At equal $\varphi$, $D_s$ of ideal TMCT is larger than
that of ideal MCT, and it vanishes at the higher critical volume fraction; the magnitudes are
not compared with the simulation data here (Sec.~\ref{sec:discussion}). Near $\varphi_c$,
$D_s\propto\epsilon^\gamma$ with $\epsilon=1-\varphi/\varphi_c$ (Sec.~\ref{sec:beta}). Figure~\ref{fig:lgam} shows
the local exponent $\dd\ln D_s/\dd\ln\epsilon$. It approaches the asymptotic value
$\gamma=1/(2a)+1/(2b)$ predicted from $\lambda$ only slowly. At $q^*$, ideal TMCT gives
$\dd\ln D_s/\dd\ln\epsilon=3.19$ at $\epsilon\simeq0.002$, consistent with an approach to
$\gamma=3.35$. In
the range $0.04\le\epsilon\le0.15$, which is typically used to fit simulation data, the
effective exponent (the slope of $\ln D_s$ between the two ends, with the value at
$\epsilon=0.04$ interpolated) is 2.60 for ideal TMCT at $q^*$
and 2.44 at $q_c\sigma=25.4$, below the asymptotic values 3.35 and 2.92. The corresponding
values for ideal MCT are 1.91 and 1.85, below the asymptotic values 2.34 and 2.28. In both theories the effective exponent in this range is therefore substantially smaller than the asymptotic one.
For comparison, the molecular-dynamics data in Fig.~\ref{fig:lgam} give local exponents
between 1.7 and 2.2 in the same range.

\begin{figure*}
\includegraphics[width=\textwidth]{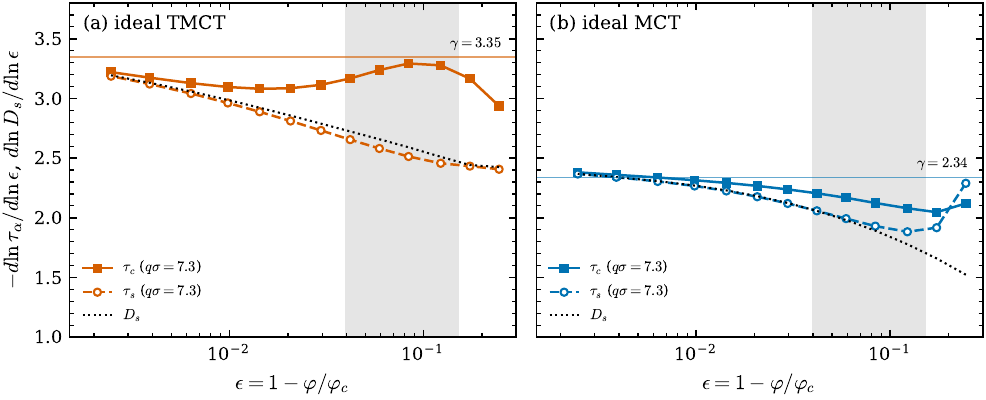}
\caption{Local exponents of the $\alpha$-relaxation times, $-\dd\ln\tau_\alpha/\dd\ln\epsilon$,
at $q\sigma=7.3$ and $q_c=q^*$ for (a) ideal TMCT and (b) ideal MCT. Squares: collective
$\tau_c$; circles: self $\tau_s$; dotted line: $\dd\ln D_s/\dd\ln\epsilon$. Horizontal lines:
asymptotic value $\gamma$. Shaded band: range typically used to fit simulation data.}
\label{fig:tau}
\end{figure*}

The $\alpha$-relaxation times, defined by $f_\alpha(q,\tau_\alpha)=e^{-1}$ at $q\sigma=7.3$,
test the same asymptotic law independently of the assumption $D_s\propto1/\tau$
(Fig.~\ref{fig:tau}). For $\epsilon\to0$ the local exponents of $\tau_c$, $\tau_s$, and $D_s$
approach $\gamma$ together: at $\epsilon\simeq0.0024$ they are 3.22, 3.19, and 3.19 for ideal
TMCT and 2.38, 2.37, and 2.37 for ideal MCT. The local exponent of $\tau_s$ follows that of
$D_s$ at all $\epsilon$, and $D_s\tau_s$ varies by less than 20\% for ideal TMCT over
$0.002\le\epsilon\le0.4$ and by less than 1\% for ideal MCT for $\epsilon\le0.05$. In the
range $0.04\le\epsilon\le0.15$, however, the effective exponent of $\tau_c$ for ideal TMCT is
3.25, much larger than those of $\tau_s$ (2.54) and $D_s$ (2.60); for ideal MCT the
corresponding values are 2.14, 1.95, and 1.91.

The difference arises from the way in which the fast, noncritical dynamics enters each
quantity. The inverse diffusion coefficient, $1/D_s=\int_0^\infty\Delta\varphi_s^{(0)}(t)\,\dd t$,
is a time integral, to which the fast dynamics contributes additively: for ideal TMCT, about
65\% of the integral comes from $t<t_0$ at $\epsilon=0.14$, 16\% at $\epsilon=0.056$, and a
negligible fraction at $\epsilon=0.004$. This nearly $\epsilon$-independent part lowers the
effective exponent of $D_s$. The relaxation time $\tau_c$, in contrast, is the time at which
$f_c$ crosses $e^{-1}$. Near the main peak, $f_q^{\infty,c}\simeq0.97$ in ideal TMCT, so that
the fast dynamics accounts for only a few percent of the decay and the crossing is
determined by the structural relaxation. Consequently, $D_s\tau_c$ increases by a factor of
3.6 for ideal TMCT and 1.6 for ideal MCT from $\epsilon=0.15$ to 0.002, whereas both theories
predict the same asymptotic exponent for $\tau_c$ and $D_s$. The self relaxation time follows
$D_s$ rather than $\tau_c$: $q^2D_s\tau_s$ at $q\sigma=7.3$ stays between 0.50 and 0.60 for ideal
TMCT and between 0.62 and 0.66 for ideal MCT for $\epsilon\le0.15$, i.e., the $\alpha$ decay of
$f_s$ at this wave number is close to a diffusive decay. In ideal MCT, where
$f_q^{\infty,c}=0.82$ at the main peak, the fast dynamics contributes more to the decay of $f_c$,
and the difference between $\tau_c$ and $D_s$ is smaller. The nonmonotonic local exponent of
$\tau_c$ for ideal TMCT reflects the small critical amplitude at the main peak
(Sec.~\ref{sec:lam}).

Although $\varphi_c$ and $\lambda$ are independent of $q_c$ for $q_c\ge q^*$, $D_s$ is not.
Table~\ref{tab:qcdep} shows $D_s$ for $q_c\ge q^*$, normalized by its value at $q^*$. In both
theories $D_s$ decreases with increasing $q_c$ at all volume fractions, including
$\varphi=0.20$, without saturation up to the largest cutoff examined. For ideal TMCT
at $\varphi=0.20$ and 0.40, $1/D_s$ is linear in $\ln q_c$ to within about 1\% for
$28\le q_c\sigma\le60$. This logarithmic dependence follows from the discontinuity of the PY
direct correlation function at contact (Appendix~\ref{app:log}): the vertex
$k^4c(k)^2S(k)$ in Eq.~\eqref{eq:m0} does not decay at large $k$, while the correlators there
relax on a time of order $(kv_{th})^{-1}$, so that each interval of $\ln k$ adds a constant
amount to the friction $\int_0^\infty\Delta\varphi_s^{(0)}(t)\,\dd t$. Since this argument
involves only large wave numbers, the dependence is not specific to ideal TMCT. Because the reduction is stronger closer to $\varphi_c$, the effective exponent also
increases with $q_c$, from 2.60 at $q^*$ to 3.40 at $q_c\sigma=60$ for ideal TMCT, and from 1.91 to
2.16 at $q_c\sigma=48$ for ideal MCT. The asymptotic exponents, fixed by $\lambda$, are unaffected. Neither the magnitude of $D_s$ nor its effective exponent is thus a
cutoff-independent prediction of the ideal theories without damping, even for $q_c\ge q^*$.

\section{Discussion}
\label{sec:discussion}

\paragraph*{Comparison with simulations.}
Molecular-dynamics simulations of hard spheres give $\varphi_c\simeq0.582$--$0.592$,
depending on polydispersity, and $D_s\propto\epsilon^{2}$ in the range
$0.04\lesssim\epsilon\lesssim0.15$ \cite{Tokuyama2003,Tokuyama2007}. At $q^*$ ideal TMCT
gives $\varphi_c=0.5824$, close to the monodisperse estimate 0.5820
\cite{TokuyamaNarumi2019}, and its effective exponent in the same range, 2.60,
is closer to 2 than its asymptotic exponent 3.35. The effective exponent depends on the
cutoff, however (2.44 at $q_c\sigma=25.4$ and 3.40 at $q_c\sigma=60$), so that this comparison
is not a test of the theory. A direct comparison of the magnitude of $D_s$ is not meaningful
at $\gamma_\alpha=0$ either: $D_s$ does not converge with $q_c$ (Table~\ref{tab:qcdep}), and $D_s$ of ideal TMCT at $q^*$ is smaller than the
simulation value by a factor 2.4 at $\varphi=0.3$ and 5 at $\varphi=0.5$. A quantitative
comparison requires the damping constants, which have been determined from the simulated
mean-square displacement in Refs.~\cite{Tokuyama2017,TokuyamaNarumi2019}; we leave this for
future work. Since $\lambda$ and the asymptotic exponents are properties of the fixed point,
they do not depend on $\gamma_\alpha$.

\paragraph*{Cutoff.}
The intrinsic cutoff $q^*\simeq1/b_c$ of ideal TMCT is the wave number beyond which density
fluctuations remain ergodic at $\varphi_c$. We have used $q_c=q^*$ because $\varphi_c$ and
the exponents of ideal TMCT do not depend on the cutoff there. $D_s$, however, depends
logarithmically on $q_c$ in both theories (Table~\ref{tab:qcdep}), so that a quantitative
prediction of $D_s$ requires a physical criterion for the cutoff. The slow convergence with $q_c$, which is
most pronounced for ideal MCT, originates from the discontinuity of $c(r)$ at contact
(Appendix~\ref{app:log}). This discontinuity is a property of hard spheres and not of the PY
approximation, so that it is not removed by using $S(q)$ from simulations of hard spheres.
For particles with a soft repulsion, $c(k)$ decays faster at large $k$, and the cutoff
dependence is expected to be weaker. In applications with
damping, the cutoff is instead determined at each $\varphi$ from the simulated $D_s$, and in
the supercooled state it satisfies $q_m\le1/b\le1/b_c$, where $q_m$ is the position of the
main peak of $S(q)$ \cite{Tokuyama2017,TokuyamaNarumi2019}. Figure~\ref{fig:cutoff} shows
that in the upper part of this range, $20\lesssim q_c\sigma\le27$, $\varphi_c$ changes by
less than 0.005, whereas $\gamma$ changes from 2.2 to 3.3. A cutoff that depends on
$\varphi$ will therefore affect not only the magnitude of $D_s$ but also its apparent
exponent. A consistent determination of $\gamma_\alpha$ and $q_c$
is left for future work.

\paragraph*{Exponent parameter.}
The difference between $\lambda_\tmct$ and $\lambda_\mct$ follows from the structure of
Eq.~\eqref{eq:tmct}: the kernel is quadratic in $f$, but the equation of motion is linear in
$K=-\ln f$. The additional contribution $\tfrac12\langle K^c\rangle_w$ is small only if $f^{\infty,c}$
is close to 1 at the wave numbers that dominate the weights $w_q$, which is not the case for
hard spheres. In single-power schematic models the additional term is compensated by $\lambda_D$, and both
theories give $\lambda=1/2$ (Sec.~\ref{sec:beta}). Our result generalizes that of G\"otze and Schilling
\cite{GotzeSchilling2015} for a schematic model to a wave-vector-dependent vertex. Reference~\cite{Tokuyama2026} expects
$\lambda_\mct>\lambda_\tmct$; for PY hard spheres we find the opposite order,
$\lambda_\tmct=0.855>\lambda_\mct=0.707$, at $q_c=q^*$. The present conclusions are
restricted to ideal TMCT. Ideal TMCT follows from full TMCT by a Markov approximation, which leads
to unphysical behavior in the $\beta$ stage, such as a negative non-Gaussian parameter
\cite{Tokuyama2017,Tokuyama2026}. The $\beta$-relaxation results of the present paper may
therefore change in full TMCT, in which $\lambda$ acquires further corrections from the
non-Markovian memory term \cite{Tokuyama2026}.

\section{Summary}

We have compared ideal TMCT and ideal MCT for PY hard spheres without damping. (i) Ideal TMCT has
an intrinsic cutoff $q^*\sigma\simeq28$, beyond which $\varphi_c=0.5824$ and $\lambda=0.855$
are independent of $q_c$. (ii) The exponent parameter of ideal TMCT differs from that of ideal MCT by a
term $\tfrac12\langle K^c\rangle_w$, which we derived analytically and confirmed by three
independent methods. (iii) Below $q^*$, the exponents depend much more strongly on $q_c$ than
$\varphi_c$ does. (iv) The asymptotic power law of $D_s$ is reached only for
$\epsilon\lesssim10^{-2}$; in the range accessible to simulations the effective exponent of
ideal TMCT at $q^*$ is 2.6, compared with $\simeq2$ from molecular dynamics. (v) Without damping,
$D_s$ does not converge with $q_c$ in either theory, and the effective exponent increases
with $q_c$ even above $q^*$. Determining the damping
constants from simulations and extending the analysis to full TMCT are left for future work.

\begin{acknowledgments}
The author is grateful to M. Tokuyama for valuable discussions on the time-convolutionless
mode-coupling theory and for continuous encouragement.
\end{acknowledgments}

\section*{Data Availability}
The data that support the findings of this article, together with the code that produced
them, are openly available \cite{KimuraCode2026}.

\section*{AI Use Disclosure}
The author used Claude (Anthropic; Claude Opus models, Claude Opus 5.5 in the final stages),
accessed through the Claude app with a code-execution environment, in October and November
2026. The AI wrote and ran the numerical code and the scripts that produce the figures,
carried out and checked the derivations of Sec.~\ref{sec:beta} and
Appendices~\ref{app:beta} and \ref{app:log}, and drafted and revised the text. The author
defined the research questions and the scope of the work, made all decisions on
interpretation and presentation, and reviewed the manuscript section by section. The results
were verified by independent cross-checks: $\lambda$ was determined by four independent
methods (Sec.~\ref{sec:lam}); the expression for $\lambda$ reproduces the schematic-model
result of Ref.~\cite{GotzeSchilling2015}; the code reproduces the critical volume fractions
of Ref.~\cite{KimuraTokuyama2016}; convergence with respect to the grid spacing and the time
discretization was tested (Sec.~\ref{sec:numerics}); and the derivation of $\lambda$ was
repeated in different variables. The author checked the derivations in
Sec.~\ref{sec:beta} (including the stability matrix and the schematic models) and in
Appendices~\ref{app:beta} and \ref{app:log} by hand, and reproduced the critical volume fractions, the exponent parameters, and the
effective exponents of $D_s$ of both theories at $q^*$ by rerunning the code. The author takes full responsibility for the content of
this article.

\appendix
\section{Derivation of the ideal TMCT $\beta$-scaling equation}
\label{app:beta}

Laplace transforming Eq.~\eqref{eq:tmct} gives exactly
\begin{equation}
\frac{1}{zK_q(z)}=zm_q(z)+\frac{z^2}{\Omega_q^2},
\end{equation}
where the last term is negligible for $t\gg\Omega^{-1}$. With $zK_q=K_q^c+z\kappa_q$,
\begin{equation}
\frac{1}{zK_q}=\frac{1}{K_q^c}-\frac{z\kappa_q}{(K_q^c)^2}+\frac{(z\kappa_q)^2}{(K_q^c)^3}+\cdots .
\end{equation}
Expanding $\mathcal{F}_q$ around the critical point in $\delta f_k=f_k-f_k^{\infty,c}$, with the control
term $\delta \mathcal{F}_q\propto\varphi-\varphi_c$,
\begin{equation}
\mathcal{F}_q=\mathcal{F}_q^c+\delta \mathcal{F}_q+\sum_kC_{qk}\delta f_k+\tfrac12\sum_{k,p}D_{qkp}\delta f_k\delta f_p,
\end{equation}
inserting Eq.~\eqref{eq:expmap}, using $\mathcal{F}_q^c=1/K_q^c$, and multiplying by $-(K_q^c)^2$, we
obtain
\begin{multline}
z\kappa_q-\sum_kA_{qk}z\kappa_k=\frac{(z\kappa_q)^2}{K_q^c}
-(K_q^c)^2z\mathcal L\Big[\tfrac12\sum_kC_{qk}f_k^{\infty,c}\kappa_k^2\\
+\tfrac12\sum_{k,p}D_{qkp}f_k^{\infty,c}f_p^{\infty,c}\kappa_k\kappa_p\Big]-(K_q^c)^2\delta \mathcal{F}_q,
\end{multline}
with $A_{qk}=A^\tmct_{qk}$ given by Eq.~\eqref{eq:stabT}. Setting $\kappa_q=e_qG+\kappa_q^{(2)}$ and
projecting onto $\hat e_q$ eliminates the left-hand side. Dividing by
$-\sum_q\hat e_qe_q^2/K_q^c$ gives Eq.~\eqref{eq:betaeq} with
\begin{equation}
\tilde\sigma=\frac{\sum_q\hat e_q(K_q^c)^2\delta \mathcal{F}_q}{\sum_q\hat e_qe_q^2/K_q^c}
\label{eq:sigt}
\end{equation}
and $\lambda$ given by Eq.~\eqref{eq:lamtmct}. The same steps applied to $zf/(1-zf)$ with
$\delta f_q=e_qG$ give
Eq.~\eqref{eq:lammct}; with $e_q\to(1-f_q^{\infty,c})^2e_q$ and $\hat e_q\to\hat e_q/(1-f_q^{\infty,c})^2$ it
reduces to the standard form \cite{Gotze1991}. Writing the derivation in terms of
$\delta f$ instead of $\kappa$, or expanding the kernel in cumulants, gives the same
$\lambda$.

\section{Logarithmic cutoff dependence of \texorpdfstring{$D_s$}{Ds}}
\label{app:log}

For PY hard spheres, $c(r)$ vanishes for $r>\sigma$ and is discontinuous at contact, with
$c(\sigma^-)=-\mu_1-6\varphi\mu_2-\tfrac12\varphi\mu_1$, where
$\mu_1=(1+2\varphi)^2/(1-\varphi)^4$ and $\mu_2=-(1+\varphi/2)^2/(1-\varphi)^4$
\cite{Wertheim1963}. An integration by parts in $c(k)=(4\pi/k)\int_0^\sigma rc(r)\sin(kr)\,\dd r$
gives, for $k\sigma\gg1$,
\begin{equation}
c(k)\simeq-\frac{4\pi\sigma c(\sigma^-)\cos(k\sigma)}{k^2},
\label{eq:clarge}
\end{equation}
so that the vertex in Eq.~\eqref{eq:m0} tends to $16\pi^2\sigma^2c(\sigma^-)^2\cos^2(k\sigma)$,
with $S(k)\to1$, and does not decay. At large $k$, the correlators relax on a time of order
$(kv_{th})^{-1}$. We write $\int_0^\infty f_c(k,t)f_s(k,t)\,\dd t=\hat\tau/(kv_{th})$ with a
dimensionless constant $\hat\tau$; for free particles, $f_c f_s=e^{-k^2v_{th}^2t^2}$, as in ideal
TMCT without memory, and $\hat\tau=\sqrt\pi/2$. Inserting this into Eq.~\eqref{eq:ds} and
averaging $\cos^2(k\sigma)$ over its oscillations gives
\begin{equation}
\frac{1}{D_s}=B\ln q_c+\mathrm{const},\qquad B=\frac43\,\frac{\rho\sigma^2c(\sigma^-)^2\hat\tau}{v_{th}},
\label{eq:logB}
\end{equation}
where the constant contains the contributions from small $k$.

For free particles, Eq.~\eqref{eq:logB} gives $B\sigma v_{th}=1.33$ at $\varphi=0.20$ and 10.0
at $\varphi=0.40$. The slopes of the numerical $1/D_s$ of ideal TMCT versus $\ln q_c$ for
$28\le q_c\sigma\le60$ are 1.63 and 33.4, respectively. The free-particle estimate thus gives
the correct order of magnitude at $\varphi=0.20$. At $\varphi=0.40$ the correlators at large
$k$ decay more slowly than those of free particles, so that $\hat\tau$ is larger, while the
logarithmic form is preserved. In ideal MCT the correlators do not decay without memory, since
$f=\cos(\Omega t)$; $\hat\tau$ is then set by the memory function, and the logarithmic form
holds only approximately.

\bibliography{refs}

\end{document}